\documentclass{article}

\usepackage{microtype}
\usepackage{algorithm2e}
\usepackage{amsmath}
\usepackage{graphicx}
\usepackage{color}
\usepackage{tabu}
\usepackage{longtable}

\title{Large-scale EM Analysis of the {\em Drosophila} Antennal
Lobe with Automatically Computed Synapse Point Clouds}

\author{Ting Zhao, Shin-ya Takemura, Gary B. Huang, \\
Jane Anne Horne, William T. Katz, \\
Kazunori Shinomiya, Louis K. Scheffer, \\
Ian A. Meinertzhagen, Patricia K. Rivlin, Stephen M. Plaza}

\begin{document}
\maketitle

\begin{abstract}
The promise of extracting connectomes and performing
useful analysis on large electron microscopy (EM) datasets has been an elusive
dream for many years.  Tracing in even the smallest
portions of neuropil requires copious human annotation,
the rate-limiting step for generating a connectome.  While
a combination of improved imaging and automatic segmentation
will lead to the analysis of increasingly large volumes, machines
still fail to reach the quality of human tracers.  Unfortunately,
small errors in image segmentation can lead to catastrophic
distortions of the connectome.

In this paper, to analyze very
large datasets, we explore different mechanisms
that are less sensitive to errors in automation.
Namely, we advocate and deploy extensive synapse detection on the
entire antennal lobe (AL) neuropil in the brain of the fruit fly
{\em Drosophila}, a region much larger than any densely annotated to date.  The resulting synapse point cloud
produced is invaluable for determining compartment boundaries in the
AL and choosing specific regions for subsequent analysis.  We
introduce our methodology in this paper for region selection
and show both manual and automatic synapse annotation results.  Finally, we note the correspondence between image datasets obtained using the synaptic marker, antibody nc82,
and our datasets enabling registration between light and
EM image modalities.
\end{abstract}

\section{Introduction}
\begin{figure}
\centering
\includegraphics[width=1.0\textwidth]{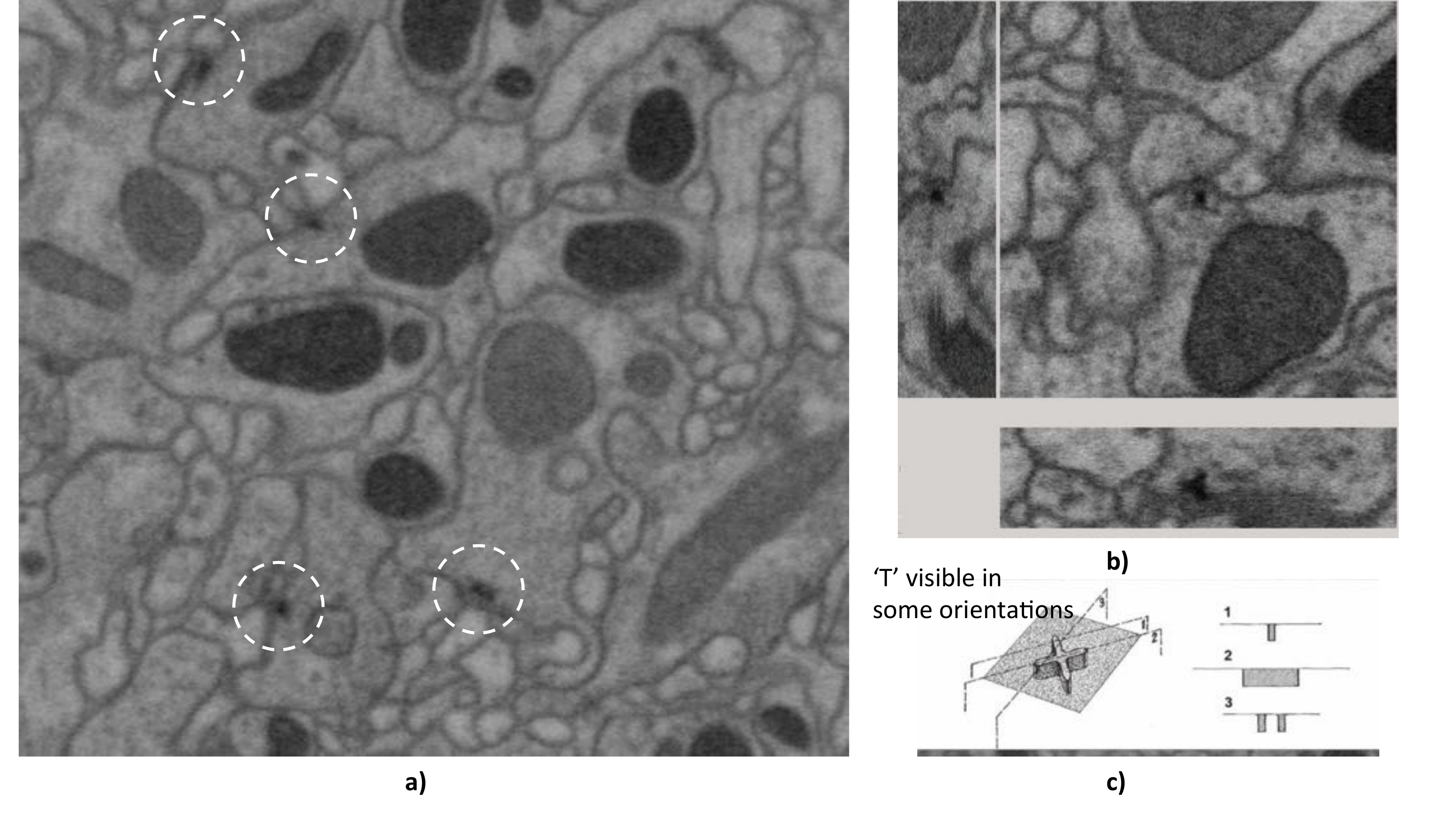}
\caption{\label{fig:synapses} {\bf Examples of synapses in EM dataset.}  a) The dashed circles highlight a few synapses in the antennal lobe of {\em Drosophila}.  They are often characterized by a T shaped structure (called a T-bar) and each T-bar has multiple post-synaptic partners.  b) The same T-bar in three orthogonal views reveals that
it has the form of a tiny table, comprising a platform surmounting a cruciform pedestal.  c)  Cartoon
depicting the structure of the T-bar (adapted from Trujillo-Cen\'{o}z, 1969 \cite{Trujillo69}).}
\end{figure}

EM data provide a definitive way to identify synapses from the apposition of
pre- and postsynaptic organelles.  Figure \ref{fig:synapses} shows
synapses in {\em Drosophila} from the locations of presynaptic
organelles called T-bars, which are T-shaped when cut in cross section.
Each synapse is clearly observed to have a presynaptic density and often a
characteristic polyad of postsynaptic partners.  Advances in machine
learning algorithms that use these characteristics \cite{huang14, kresuk} could enable large-scale,
automated identification of synaptic contacts, which could eliminate the need for
time-consuming manual annotation \cite{plaza14synapse}.  Note that while
synapse prediction promises to be an important contributor to connectomics,
the task of tracing neurons is still laborious \cite{Nature13, Winfried13}.

In this paper, we introduce high-level applications of synapse prediction in
EM that require minimal manual effort.  In particular, we exploit the fact that  
identifying synapses in EM can be used to generate a {\em synapse point
cloud} that is the high-resolution
analogue to the nc82 label of Bruchpilot (Brp), a synaptic protein \cite{Wagh06}.
Synapse prediction reveals high-level neuropil structure not always
evident from inspection of lower magnification EM data.
Synapse prediction from EM data offers moreover advantages
over nc82 immunolabeled data, because the accuracy of synapse density across
neuropil regions can be verified by inspecting the high resolution EM
data.
%Second, the ease of creating training data for synapse prediction \cite{huang14}
%suggests greater flexibility in classifying interesting synaptic phenomenon without
%requiring specific genetic label sites.

Our primary contribution
lies in using synapse point clouds to instantly identify the boundaries
of a neuropil in EM with great accuracy.  This contribution leads to three applications:

\begin{enumerate}
\item Defining accurately regions of interest (ROIs).  Having accurate ROIs is crucial to reconstruct connectomes because an imprecisely defined ROI leads either to unnecessary, costly manual effort or an incomplete result.
%We introduce software to quickly identify these regions.
\item Registration of EM synapse point cloud to light microscopy (LM) data to facilitate neuropil
identification and exploit information annotated in the LM datasets especially in
standard brain atlases \cite{flylight}.
%We introduce semi-automated clustering strategies to facilitate these comparisons.
\item Verifiable statistics of synapse packing
density across different brain regions not limited
to the outputs of particular cell types as in \cite{mosca14}.
\end{enumerate}

\begin{figure}
\centering
\includegraphics[width=1.0\textwidth]{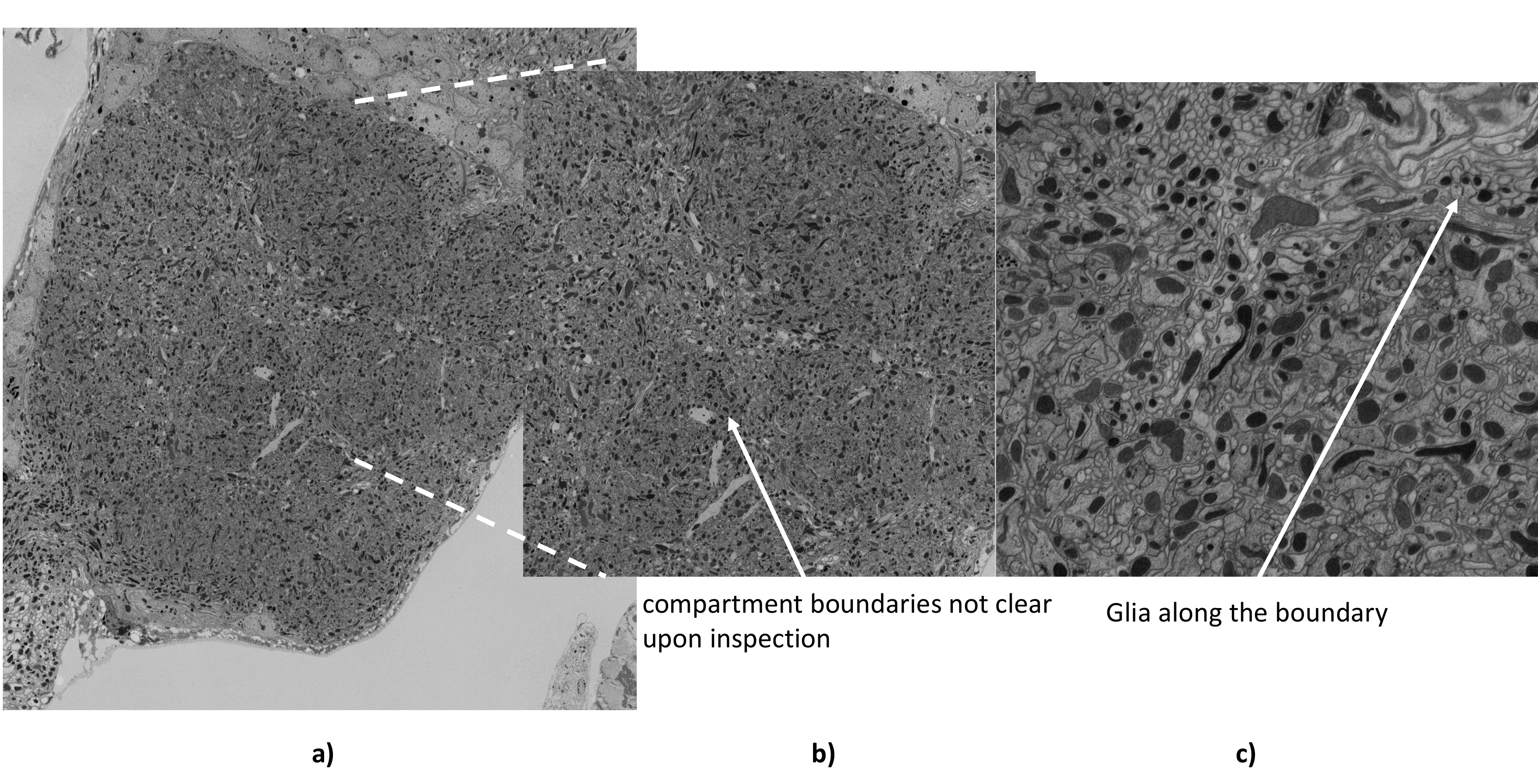}
\caption{\label{fig:alunclear} {\bf Precise neuropil boundaries are difficult to discern from EM data alone.}  a) The antennal lobe reveals some subtle differences in neuropil regions at lower magnification. b) With enlarged magnification, some of the boundaries become evident, while others remain unclear.  c)  Closer inspection reveals glia that help define the boundaries between glomeruli.}
\end{figure}

To test our methods, we have studied the organization
and synapse density of the {\em Drosophila} antennal lobe (AL).  Several light
microscopic have previously classified the glomeruli
\cite{Kondoh03, laissue99, couto05, grabe14}, neuropil compartments, in the AL, but the
exact number has been uncertain.  Glomeruli
vary greatly in size depending on sex, sub-species, experience, and so on.  This variation
has confounded efforts to demarcate and classify the glomeruli precisely.  While ideal comparisons control
for genotype, experience level, etc, the size, shape, and locations
of glomeruli can vary greatly and non-uniformly between {\em in vivo}
and {\em in vitro} preparations, as a result of shrinkage and other disruptions from dissection \cite{grabe14}.
Inspection of the EM data alone is insufficient to discern the
boundaries between neighboring glomeruli (Figure \ref{fig:alunclear}).  While the glia on the boundaries
are often visible, it is often difficult to
untangle these accurately from the rest of the neuropil, and as a result their locations
fail to arbitrate the boarders between glomeruli.

By contrast, application of our synapse prediction in the AL provides synapse point clouds of around 500,000
points that clearly reveal neuropil boundaries.  We show how these clear boundaries
lead to precise ROIs that could significantly reduce manual annotation effort.
Furthermore, we successfully clustered this point cloud and were able to classify
prominent glomeruli consistent with LM data.  A reasonable registration
between LM and EM synapse point clouds in such a variable neuropil provides some evidence
that these techniques will generalize to other regions in the {\em Drosophila} brain.
Finally, we extract synapse counts
in different glomeruli that have been statistically validated by manually
consulting the source EM data.  Initial findings are compared against those
from the previous literature and suggest potential additional analyses.

Specific contributions in this paper include:
\begin{enumerate}
\item  Application of synapse predictions over a large
region with a small amount of training data.  The quality of these
predictions suggests that object classification can be generalized
across a large brain compartment.
\item Methods to efficiently define a region of interest
using the synapse point cloud.
\item Clustering strategies that aid in the identification
of AL glomeruli.
\item Visualization techniques for point cloud data that allow these to be
registered with data from light microscopy.
\end{enumerate}

We will first introduce our new methods.  Then we will show results from applying
synapse prediction to the AL, where this information is used to identify
neuropil compartments and help define a ROI to better concentrate proofreading efforts.
We conclude by discussing applications of these methods.

%% add picture of tbar, cannot easily spot in nc82, recent work still leaves a lot of unanswered questions, EM
%% was always slow to analyze
%% add picture of AL grayscale -- hard to see anything

\section{Methods}
\begin{figure}
\centering
\includegraphics[width=1.0\textwidth]{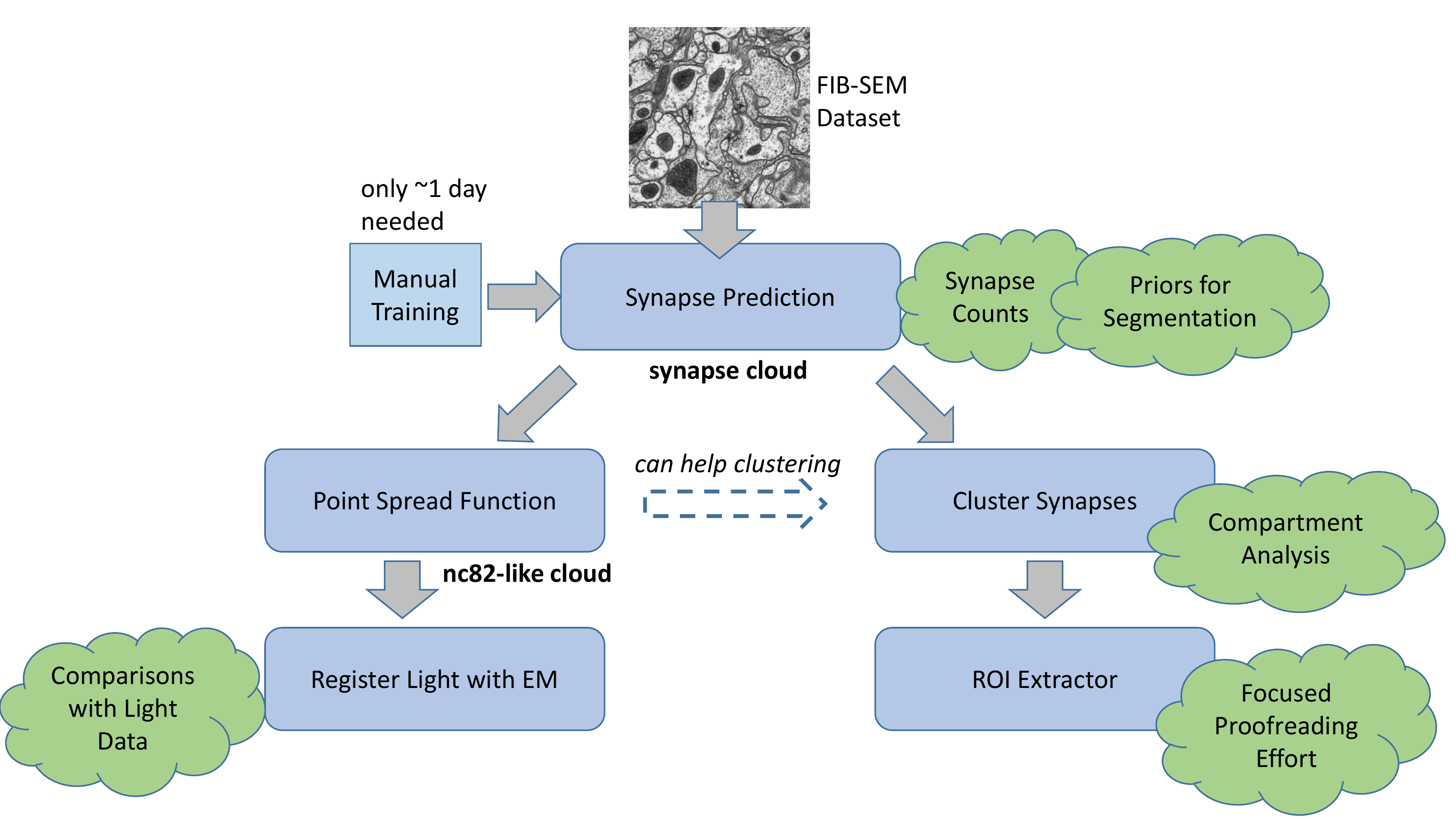}
\caption{\label{fig:flow} {\bf ROI and Synapse Methodology.}  The light blue boxes show the
parts of our methods.  The green clouds indicate potential applications explored in this paper.}
\end{figure}

Figure \ref{fig:flow} provides an overview of the methods introduced in this paper and the applications of those
methods.  In particular, it highlights our strategy for exploiting scalable, automatic synapse prediction.

We use images acquired by FIB-SEM techniques \cite{fib} to produce a large isotropic dataset.  The isotropic resolution
improves image alignment and, generally, the effectiveness of machine learning algorithms \cite{plaza14focused},
which is critical when analyzing very large datasets \cite{plaza14}.  We
then produce synapse predictions following the procedure outlined in \cite{huang14}.  To apply this approach
to a very large dataset, both the training required for the classifier and the prediction
algorithm must scale to handle very large datasets.  Fortunately, the approach in \cite{huang14} has shown itself
to be effective with only a small amount of training.  We later show results in which the prediction generalizes
over the entire antennal lobe in {\em Drosophila}.  However, in practice, one might need to generate training
for each distinct neuropil region.  The main computational bottleneck in \cite{huang14} lies in computing the
probability that each voxel belongs to a synapse.  This step is distributed across a compute cluster
with each local worker processing a small spatial subvolume of the entire region.

The application of synapse prediction produces a {\em synapse point cloud}.  Unlike algorithms that
automatically extract neuron shapes from EM, in which small automatic errors can exert
considerable
impact on the connectome \cite{plaza14focused}, we conjecture that small inaccuracies in the synapse point cloud
will have only a minimal effect on our ability to analyze high-level trends.  Given the resolution of EM data,
these predictions can be sampled and accurate synapse counts for different regions can be provided, which
is generally difficult using nc82 immunolabeling.  We also suggest the potential for using these predictions
as hints for image segmentation algorithms.

\vspace{2mm}
\noindent {\bf Synapse Density Maps.}  We apply a Gaussian point spread function (PSF) to the synapse point cloud to generate a smooth map of synapse density. If the synapse cloud contains $N$ points and the center of the $i$th point is $\mathbf{c_i}$, the density map is defined as:

\begin{equation}
D(\mathbf{x}) = t \sum_{i=1}^N e^{-\frac{(\mathbf{x} - \mathbf{c_i})^2}{2}}
\end{equation}

\noindent where $\mathbf{x}$ is a point in the 3D space and $t$ is the normalization factor, which scales the density to cover the full value range for an image of a given bit depth.

The density map has several applications.  For example, the
density map resembles the lower 
resolution data from nc82 light microscopy.  Given this, we can register light and
EM datasets from similar neuropil regions using the techniques introduced in \cite{aso14}.
Because nc82 and EM preparations are different, there will be neuropil variations.  Despite these variations, our results
show that we are still able to register prominent neuropil landmarks.  Such registration could guide
neuronal tracing in EM data by using high-level information from light microscopy, although we do not pursue that application in this paper.

The density map also allows us to cluster the image into different neuropil regions through image segmentation.  This is done by resampling the map to produce a 3D digital image that can be directly fed into any image segmentation method.  We perform segmentation using a semi-automated approach.
While clustering with unsupervised approaches like hierarchical-based clustering
is sometimes possible when applied directly on
a synapse point cloud, we achieved a better clustering result
by manually adding seeds to disjoint compartments visible
in the density map.  Seeded watershed \cite{watershed} is then applied on the density map
to achieve  the final clustering (thus avoiding the need to laboriously outline the exact boundaries in three dimensions).  In
many cases, subtle changes and discontinuities in the point cloud are picked up by the automated algorithms, allowing one to
see compartment boundaries clearly.  This information can be used to define neuropil regions, as we show later for the antennal
lobe.

To obtain good segmentation results, the Gaussian PSF should reveal
the gaps between the compartments while making each compartment as smooth as possible. According to the light microscope data, the minimal gap size between neighboring compartments is about $2\mu m$, which means the maximal value of $\sigma$ should be similar. In our implementation, density map resampling resulted in a 3D image with the voxel size $0.16\mu m$ along each dimension, which means $\sigma \le 12.5$ in pixel units. We tested two value $\sigma = 5$ and $\sigma = 10$ and found that $\sigma = 10$  performed better for the seeded watershed algorithm used.

\vspace{2mm}
\noindent {\bf Defining ROIs with Point Clouds.}  These neuropil clusters can be invaluable for defining ROIs for EM proofreading and used to validate previous boundaries as determined by LM data.  For instance, it is often desirable
to answer questions like: What are all of the connections within
a specific brain compartment?  We therefore need ways to annotate
which portions in the neuropil are in this region and which not.  Because
the compartments cannot be precisely defined by a simple bounding box,
in general, we need a different approach to define the region.  To this
end, we introduce a set of algorithms (described below)
in the package Neutu \cite{neutu} and DVID \cite{dvid}.

\begin{figure}
\centering
\includegraphics[width=1.0\textwidth]{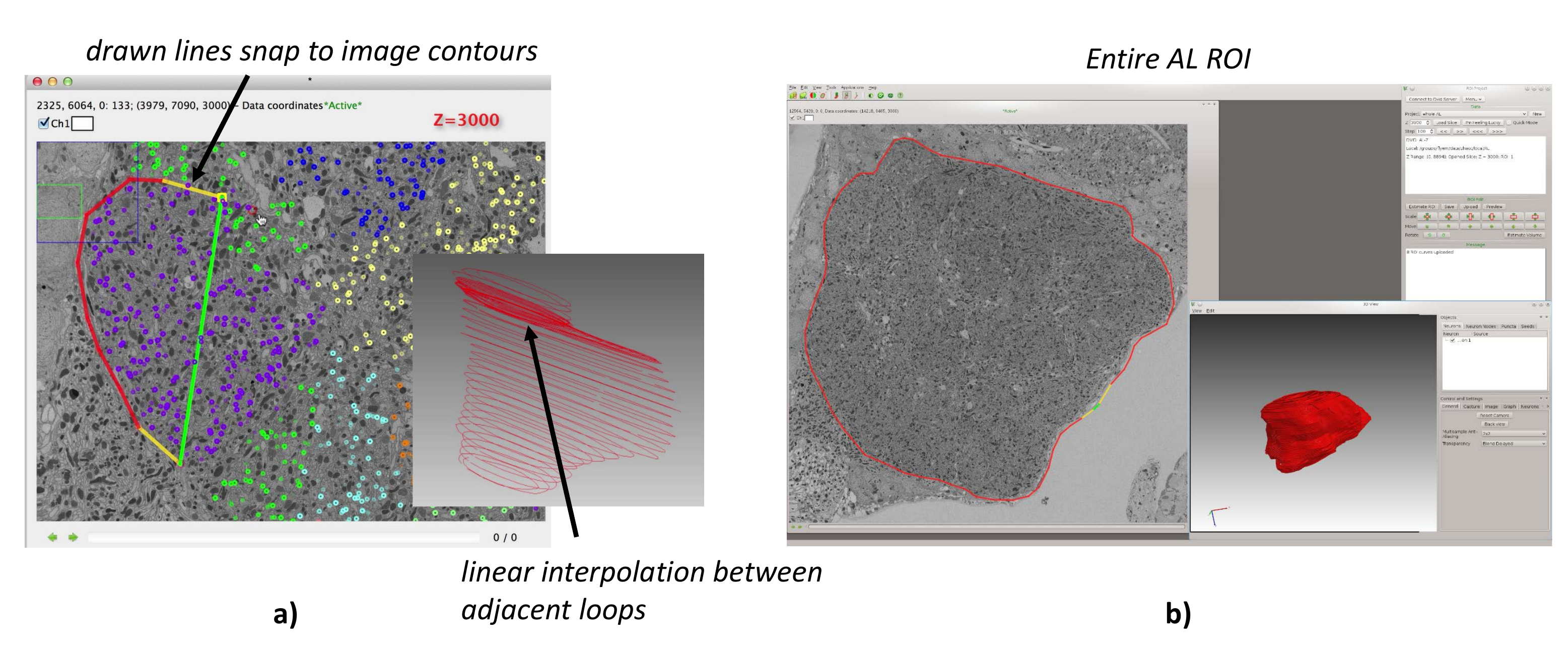}
\caption{\label{fig:roiflow} {\bf Extracting an ROI from an EM dataset.}  a) Tracing algorithm implemented in Neutu allows a proofreader to select a large region efficiently using clustered synapse predictions as a guide.  b)  Shows the
ROI for the entire AL (around $300 \cdot 10^3 \mu$m$^3$).}
\end{figure}

DVID defines an ROI at the granularity of small blocks (32x32x32 pixels).  Each ROI consists of a list of coordinates to these blocks.  In this manner, we can more concisely define an arbitrary region with only a small sacrifice in precision.  The tool Neutu enables users
to quickly define an ROI by drawing {\em loops} in different image planes.  The user can draw the loop as a series of line segments that
will snap to the contours, as shown in Figure \ref{fig:roiflow}a.  The contour is chosen as a shortest path through the grayscale data
between the two endpoints of the line segment.
The user does not have to draw the ROI completely.  Neutu will interpolate between loops drawn on non-consecutive planes.  Using this approach, we can precisely and quickly
define neuropils.  An ROI encompassing the entire AL was quickly produced and is shown
in Figure \ref{fig:roiflow}b.

As we saw in Figure \ref{fig:alunclear}, it is often hard to determine
the boundary of the neuropil from the EM data alone.  The synapse
clusters provide hints for defining compartment boundaries.  In Section \ref{sec:roi}, we show how precise definition of the region produces significant reduction of manual EM proofreading effort, the primary bottleneck to scaling connectome reconstruction.

\section{Empirical Evaluation}
To test our methodology, we analyzed a dataset containing the AL of the
3-day old adult, female {\em Drosophila melanogaster}.  The sample was prepared from a 250$\mu$m frontal head slice of a fly obtained by crossing homozygous w1118 with CS wild type, prefixed for 20 min in 2.5\% each of paraformaldehyde and glutaraldehyde in 0.1 cacodylate buffer, then high-pressure frozen and freeze substituted in 1\% OsO4 and 0.2\% uranyl acetate.
A volume roughly containing the entire AL was imaged from 2nm milling
depths using FIB and 8nm
resolution using SEM.  For consecutive image planes were averaged to yield 8nm resolution in the z-axis. Alignment was then performed on the dataset to produce
a 3D image stack with 8x8x8 nm resolution.

\subsection{Synapse Prediction}

To predict synapses in the AL, we first manually annotated a subvolume of the AL
roughly encompassing one glomerulus.  This region was split into training and validation, in
order to train a classifier for automatic T-bar detection, and to set
the threshold for the classifier to yield approximately 0.9 recall ({\em i.e.}, so that
$90\%$ of real T-bars
were found) corresponding to a precision of 0.8 ({\em i.e.}, $80\%$ of the T-bars found
were correct) on the validation data.

In our implementation, we made use of a shared cluster resource in which
up to 2500 cores were available (with variations depending on load
from other users).
The trained classifier was then applied to the entire AL,
which was about 680 gigavoxels in size. Inference
over this full volume took 11 days, and produced approximately $520
\cdot 10^3$ T-bar detections.
%% number of predictions based on region where top Z is removed

\begin{figure}[hbt]
\centering
\includegraphics[width=1.0\textwidth]{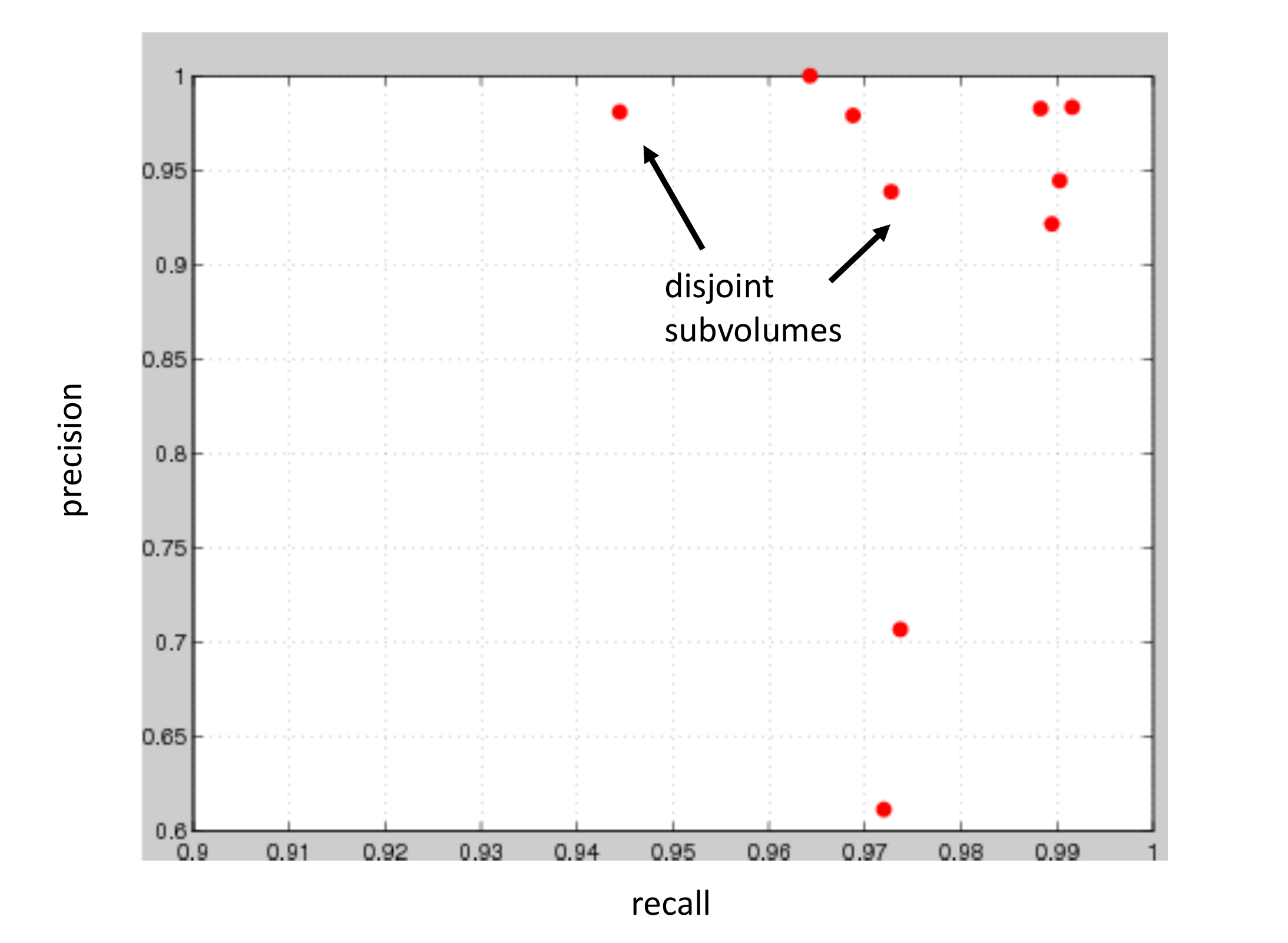}
\caption{\label{fig:generalizable} {\bf Generalizability of the synapse classifier over the AL.}  Automated T-bar predictions within 10 randomly chosen subvolumes were manually inspected.  The precision and recall is over 90 percent for most subvolumes.  There are two subvolumes with lower precision ($< 0.9$).}
\end{figure}

\begin{figure}[htbp]
\centering
\includegraphics[width=0.8\textwidth]{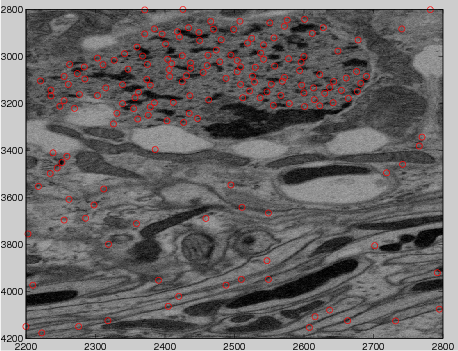}
\caption{\label{fig:badpred} {\bf Examples of neuropil with over-prediction.}  Dark granules in neuronal soma
caused several false positives in the predictor.  Normalizing the image contrast improves the quality of the prediction.}
\end{figure}

To assess how well the classifier generalized across the
entire image region, predictions within a small number of
randomly selected subvolumes, spread across the whole ROI, were manually reviewed.
Figure \ref{fig:generalizable} indicates precision and recall are over 90\%
for most subvolumes.  However, some outliers exist, which are a result
of darker regions in the image volume, as shown in Figure \ref{fig:badpred}.
For example, dark granules in a soma result in several mis-predictions.  A negative
result is not surprising since the somata exist at the periphery
of the neuropil in the fly brain and were not included in our training dataset.  

\begin{figure}[htbp]
\centering
\includegraphics[width=1.0\textwidth]{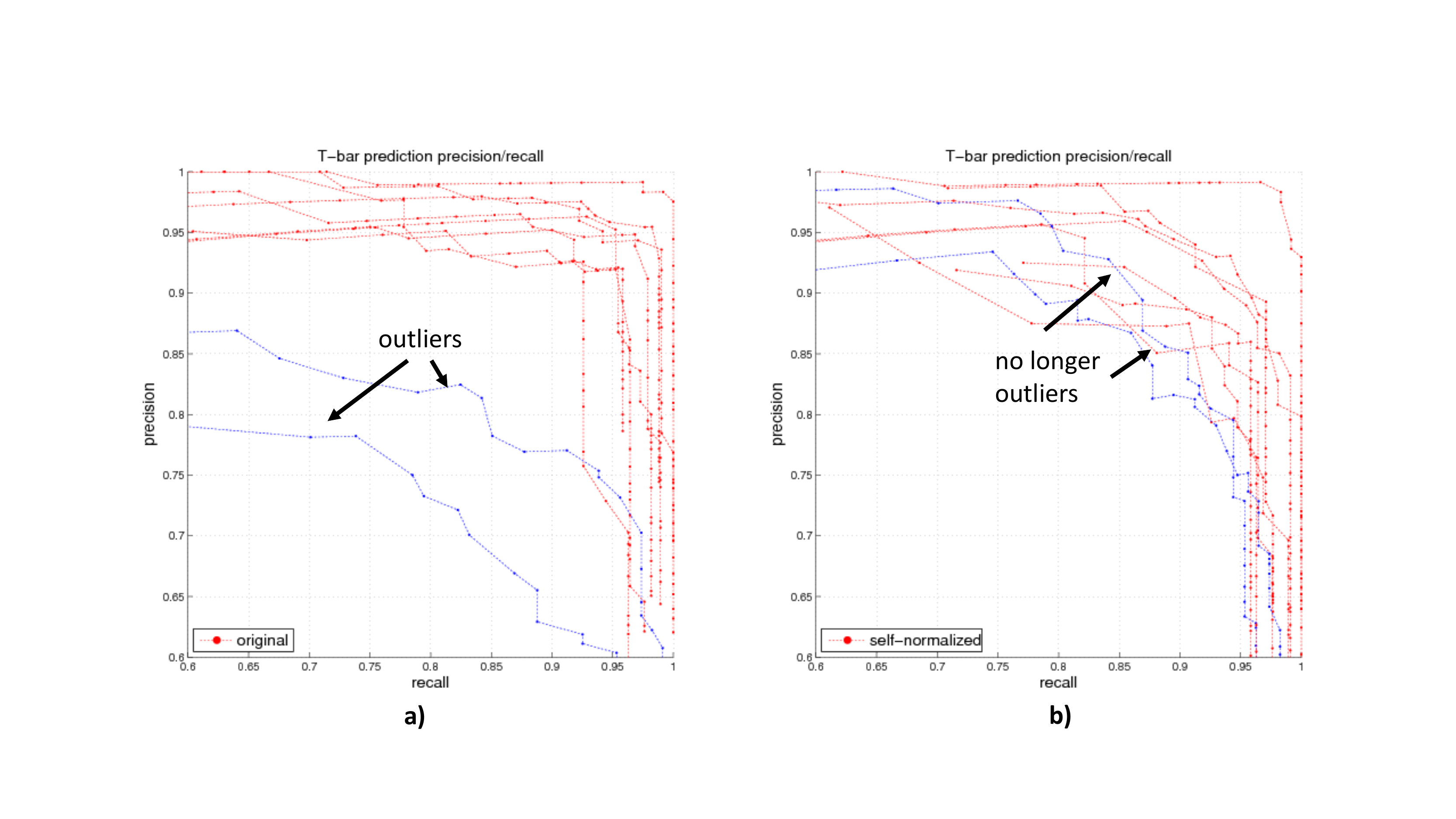}
\caption{\label{fig:normalize} {\bf Improved consistency in prediction after normalization.}  a) The precision and recall over 10 subvolumes without normalization leads to two outliers.  b)  These outliers are removed when normalizing each subvolume to 0 mean and unit standard deviation.}
\end{figure}

These generalization errors can potentially be resolved using multiple approaches.  One simple technique is
to locally normalize each subvolume, such that the subvolume has 0 mean
and unit standard deviation, as a means of compensating for regions where the image has darker intensity. 
As shown in Figure \ref{fig:normalize}, the low precision outlier substacks
are significantly improved by this normalization, with only a slight average degradation to the rest of the samples.  

Alternatively, additional training samples could be collected from regions where the original classifier
performed poorly, and used to either re-train the classifier, or train a new classifier specific to regions 
where the original classifier did not generalize well.

While the previous analyses show a good overall prediction accuracy
and that the variance in accuracy
between random regions is reasonably consistent, subtle biases between
regions could undermine particular analyses.  We will explore this shortly
in the context of estimating the synaptic density across the AL glomeruli.

\subsection{Synapse Cloud and Registration}
We showed that synapse prediction can be done with a small amount of training in a
manner that generalizes well statistically across the entire AL.  
As a subsequent evaluation of the robustness of our result, we compared it
against preparations in which nc82 antibody was used to label
the Brp protein at pre-synaptic T-bars.

\begin{figure}
\centering
\includegraphics[width=1.0\textwidth]{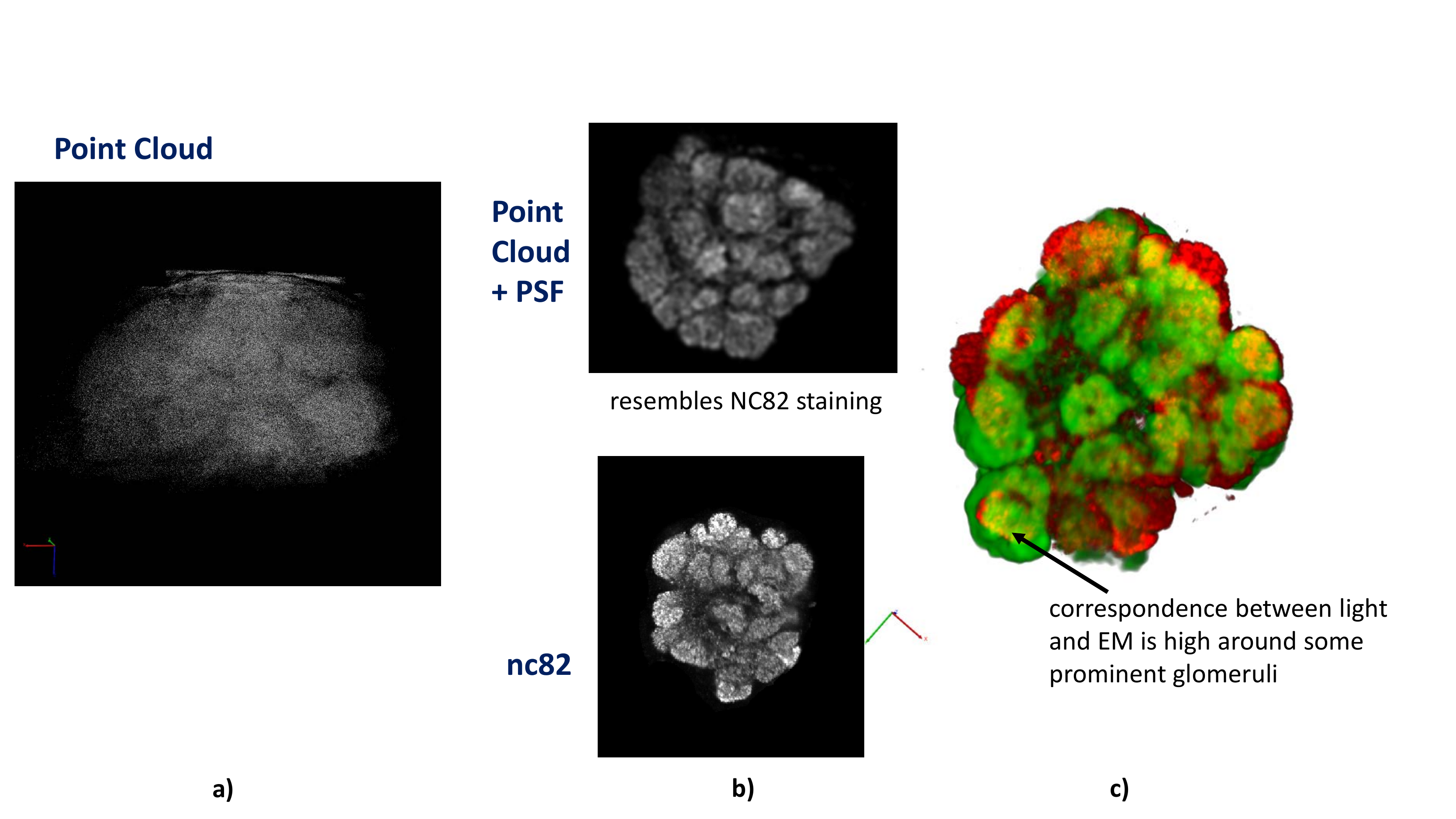}
\caption{\label{fig:register} {\bf From synapse point cloud to EM/light registration.}  a) Synapse cloud contains over 500,000
synaptic points.  It is difficult to determine the biological compartments.  b) A Gaussian point-spread function is applied producing data that looks similar to nc82 immuno-images, revealing
glomeruli compartments in the AL.  c) Registration of EM and LM volumes shows good correspondence for some of the more prominent
compartments.}
\end{figure}

Figure \ref{fig:register}a shows a synapse cloud containing over 500,000 synaptic points.  After applying a Gaussian
point spread function over the data in Figure \ref{fig:register}b
we can discern some of the AL compartments, noting
qualitative similarities to confocal images of an nc82 dataset.  Because of differences in the protocols used to prepare images between our sample and the nc82 dataset,
we expect variation in glomeruli size and
shape.  While the inability to register the two datasets would not necessary invalidate our techniques,
it is significant that even a rough registration was possible, as shown in \ref{fig:register}c, suggesting the
biological relevance of our EM and machine
learning approach to the problem.  Note that some of the prominent glomeruli regions are common in both datasets.

\subsection{Application 1: Defining Glomeruli and Analyzing Synapse Density in the AL}
\begin{figure}
\centering
\includegraphics[width=1.0\textwidth]{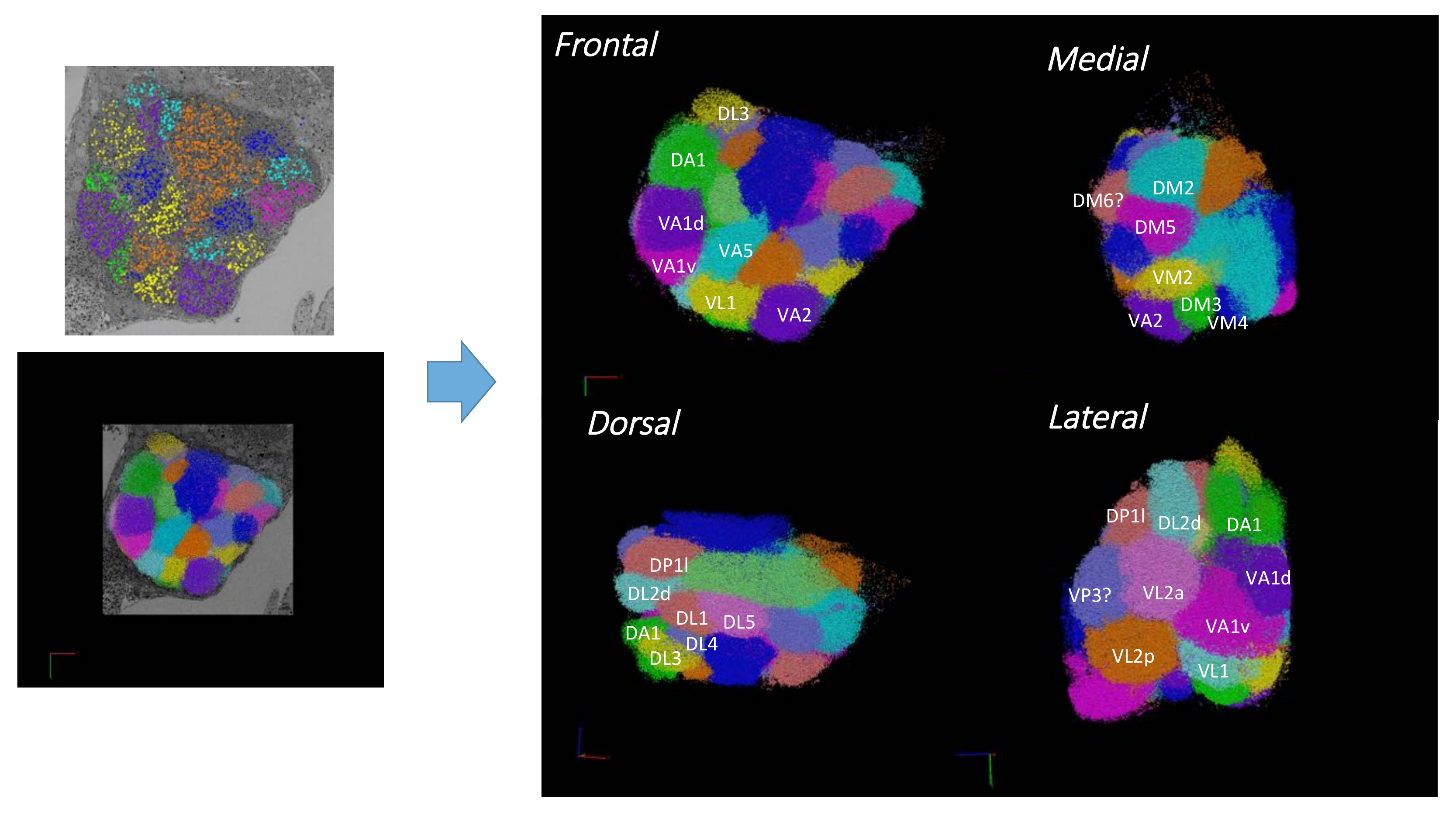}
\caption{\label{fig:cluster} {\bf Semi-automatic synapse cloud clustering and identification of several glomeruli.}  Around 60 regions
are first conservatively segmented using the synapse point cloud and seeded watershed with manually placed seeds.  The
clusters are then manually refined to produce 50
regions. 26 of these regions are given names consistent with those previously
reported in the literature.}
\end{figure}

Figure \ref{fig:cluster} shows the result of semi-automatic clustering of the synapse point cloud followed by some manual revision of the clusters.  We identify 49 compartments, close to the 54 identified in a previous study \cite{grabe14}.
As noted, the exact number and size of AL glomeruli are both
difficult to characterize definitively.
This is due in part to intraspecies variability and sensitivity to image preparation procedures.
We do not aim to clarify this situation.  Instead, we focus on whether the synapse point cloud
can be used to define glomerular boundaries in general.
Figure \ref{fig:cluster} shows labels for
some of the more prominent glomeruli.  The proper registration of this
dataset to a standard brain derived from nc82 samples makes the task of labeling more straightforward.
Notice that several prominent glomeruli are clearly identified.  The feasibility of this
task confirms that synapse clouds can be used for such analysis.

We also consider whether automatic synapse prediction permits analysis
of variations of synaptic density across glomeruli.
Making a strong claim about synaptic variation in the presence of
approximate synapse prediction across glomeruli with ambiguous
boundaries may not be possible.  Also, comparing synaptic density between
our sample and others is further complicated by the non-uniform shrinkage that occurs
in the AL due to image preparation \cite{grabe14}.
To improve the analysis, we restrict our attention
to glomeruli with boundaries and in which the
correspondence to previous literature is strongest.
Future work could entail generating predictions over multiple EM datasets, which would greatly
improve the robustness of our result.

Table \ref{tab:segstats} show the size
and synapse density for all compartments of the AL.  The labeled glomeruli are indicated with appropriate
names.
The high variation in glomerular size corroborates previous anatomical studies from
LM.  The density differences
are more interesting.  Synapse density range from under 2 per cubic micron to over 3 per cubic micron.
An overall density of 2.4 synapses per cubic micron is predicted (perhaps slightly more in practice
depending on the recall as discussed in the next paragraph).  Do these density variations reflect
reality or are they the outcome of expected variation in the prediction, along with non-uniform size
changes of the glomeruli that result from image preparation?

\begin{center}
\begin{longtabu}{|r|r|r|r|r|}
%\begin{table}[htb]
%\centering
%\begin{tabular}{|r|r|r|}\hline
%\#synapses & cubic microns & \#synapses/ \\ 
% & & cubic microns \\ \hline\hline

\caption{
{\bf Density of predicted synapses in different segmented compartments.}
The density of synapses range from 1.5 to 3.3 synapse per cubic micron.
The target glomerulus for manual annotation was VA1v.  The last column adjusts
the synapse density based on manual spot checking.} \label{tab:segstats} \\

\hline \multicolumn{1}{|c|}{\textbf{glomerulus}} & \multicolumn{1}{c|}{\textbf{\#synapses}} & \multicolumn{1}{c|}{\textbf{cubic microns}} & \multicolumn{1}{c|}{\textbf{\#synapses/}} & \multicolumn{1}{c|}{\textbf{adjusted}} \\ 
& & & \multicolumn{1}{c|}{\textbf{cubic microns}} & \\ \hline 
\endfirsthead

\multicolumn{5}{c}%
{{\bfseries \tablename\ \thetable{} -- continued from previous page}} \\

\hline \multicolumn{1}{|c|}{\textbf{glomerulus}} & \multicolumn{1}{c|}{\textbf{\#synapses}} & \multicolumn{1}{c|}{\textbf{cubic microns}} & \multicolumn{1}{c|}{\textbf{\#synapses/}}  & \multicolumn{1}{c|}{\textbf{adjusted}} \\ 
& & & \multicolumn{1}{c|}{\textbf{cubic microns}} & \\ \hline
\endhead

\hline \multicolumn{5}{|r|}{{Continued on next page}} \\ \hline
\endfoot

\hline
\endlastfoot

unlabeled & 804 &  526 &  1.53 & unchecked \\
unlabeled & 1002 &  546 &  1.84 & unchecked \\
unlabeled & 1676 &  748 &  2.24 & unchecked \\
unlabeled & 2031 &  955 &  2.13 & unchecked \\
DA3? & 2230 &  976 &  2.28 & unchecked \\
unlabeled & 1816 &  1012 &  1.79 & unchecked \\
VL1 &  9792 &  3614.04 &  2.71 & unchecked \\
unlabeled & 5111 &  1645 &  3.11 & unchecked \\
unlabeled & 2641 &  1655 &  1.60 & unchecked \\
DL3 &  2984 &  1694 &  1.76 & unchecked \\
unlabeled & 3796 &  1872 &  2.03 & unchecked \\
unlabeled & 4448 &  1872 &  2.38 & unchecked \\
VA4 & 5068 &  2006 &  2.53 & unchecked \\
VM3 &  5232 &  2202 &  2.38 & unchecked \\
unlabeled & 5904 &  2388 &  2.47 & unchecked \\
DM3 &  7615 &  2646 &  2.88 & unchecked \\
unlabeled & 6159 &  2648 &  2.33 & 2.64 \\
VM2 &  5814 &  2659 &  2.19 & unchecked \\
unlabeled & 6740 &  2763 &  2.44 & unchecked \\
unlabeled & 6341 &  2770 &  2.29 & 2.50 \\
unlabeled & 9969 &  3004 &  3.32 & 2.77 \\
DL5? &  8851 &  3174 &  2.79 & unchecked \\
unlabeled & 6322 &  3181 &  1.99 & unchecked \\
DM6 &  7704 &  3245 &  2.37 & unchecked \\
VA3? & 9224 &  3461 &  2.67 & unchecked \\
DL1 &  10165 & 3541 &  2.87 & unchecked \\
unlabeled & 8031 &  3712 &  2.16 & unchecked \\
DM5 &  8856 &  3995 &  2.22 & 2.69 \\
unlabeled & 11100 & 4107 &  2.70 & unchecked \\
unlabeled & 9443 &  4215 &  2.24 & 2.60 \\
unlabeled & 9683 &  4335 &  2.23 & unchecked \\
unlabeled & 8966 &  4395 &  2.04 & unchecked \\
VP3? & 12002 & 4508 &  2.66 & unchecked \\
unlabeled & 13607 & 5162 &  2.64 & 2.59 \\
VA2 & 13548 & 5399 &  2.51 & 3.01 \\
VA1d & 11669 & 5652 &  2.06 & unchecked \\
DA1 & 8778 &  5803 &  1.51 & 4.18 \\
VA5? & 18009 & 5826 &  3.09 & unchecked \\
VL2a &  16323 & 5987 &  2.73 & 3.15 \\
DP1L & 18369 & 6066 &  3.03 & unchecked \\
DM2 &  15729 & 6714 &  2.34 & unchecked \\
DL2d & 18651 & 6719 &  2.78 & unchecked \\
VL2p & 17370 & 6720 &  2.58 & unchecked \\
{\bf VA1v} & 16513 & 7021 &  2.35 & 2.20 \\
V &  20132 & 7257 &  2.77 & unchecked \\
unlabeled & 20743 & 7364 &  2.82 & unchecked \\
unlabeled & 22767 & 9452 &  2.41 & unchecked \\
unlabeled & 30951 & 16855 & 1.84 & unchecked \\
VM4 &  49165 & 25510 & 1.93 & unchecked \\
\hline
{\bf totals} &  {\bf 519844} & {\bf 219577} & {\bf 2.4} & - \\
%\hline \end{tabular}
%\caption{
%\label{tab:segstats}
%{\bf Density of synapses across segmented compartments.}}
%\end{table}
\end{longtabu}
\end{center}

We first aim to better understand the variation that results from predictor inaccuracy.
While Figure \ref{fig:generalizable} shows consistency and generalizability of the predictor, 
through manual verification of predictions in randomly sampled regions, bias is still possible.  For example,
one glomerulus could have different contrast or neuropil structures
than other glomeruli, confounding comparisons between different glomeruli.
To test for this possibility, we manually identified synapse annotations for
$10$ small subvolumes (64 $\mu$m$^3$ in size) in the center
of distinct, well-defined glomeruli that exhibit varying synaptic densities.
The hope was
to observe consistent precision / recall.  The results in Table \ref{tab:glomverify}
show that predictions over most subvolumes find
70 - 90\% of real synapses (recall) and that 70 - 90\%
of all predictions are real synapses (precision).  However, two substacks (shown in bold)
are significant outliers.  The low precision for the second subvolume appears to be the result
of a dark image as discussed previously.  {\em DA1} misses several synapses.  This
appears to result from a very low-contrast image.  We use these precision / recall numbers
to update the corresponding synapse density estimate in Table \ref{tab:segstats}.  While the variation,
even among the regions with more consistent prediction, is too large to make definitive
claims, a preponderance of glomeruli with 2.2 to 2.8 synapses per $\mu$m$^3$ is evident.
Interestingly, the manually examined subvolumes contain significant variance in identified
synapses, but insofar as
the subvolumes are only approximately in the center of each glomerulus, the variation may
result from differences within a glomerulus.  It may also reflect non-uniform volume
differences in the AL.

\begin{table}[htb]
\centering
\caption{
\label{tab:glomverify}
{\bf Accuracy of predictor compared to 10 subvolumes in the center of 10 different glomeruli.}  The predictor performs similarly for all but two of the substacks highlighted below.}
\begin{tabular}{|r|r|r|}\hline
glomerulus	& recall &	prec \\
\hline \hline
unlabeled	& 0.78 & 0.85 \\
{\bf unlabeled}	& 0.81 &	0.68 \\
DM5	& 0.73	& 0.88 \\
VA2 &	0.70 &	0.84 \\
VL2a &	0.70 &	0.81 \\
unlabeled &	0.77 &	0.89 \\
VA1v &	0.88 &	0.82 \\
{\bf DA1} &	0.29 &	0.79 \\
unlabeled &	0.88 &	0.86 \\
unlabeled &	0.79 &	0.89 \\
\hline \end{tabular}
\end{table}

In future work, we will try to better understand the density of the synapses across
glomeruli compared with those previously reported by normalizing our data
to glomeruli sizes in published {\em in vivo} studies.
This will reduce the impact of distortion caused by image preparation and
enable more accurate comparisons of synapse counts.

\subsection{Application 2: Reducing EM Proofreading Effort by Selecting the Region of Interest}
\label{sec:roi}
We now show how synapse clouds can reduce proofreading work.  We originally tried
to analyze the connectome within a glomerulus.  To determine the connectome, we first
annotated all of the synapses in a region believed to contain an entire glomeruli.  The initial boundary was based on what appeared to be glia and was known to be an approximation.  It was unclear how accurately located this initial ROI would be until we started annotating the synapses.

\begin{figure}
\centering
\includegraphics[width=1.0\textwidth]{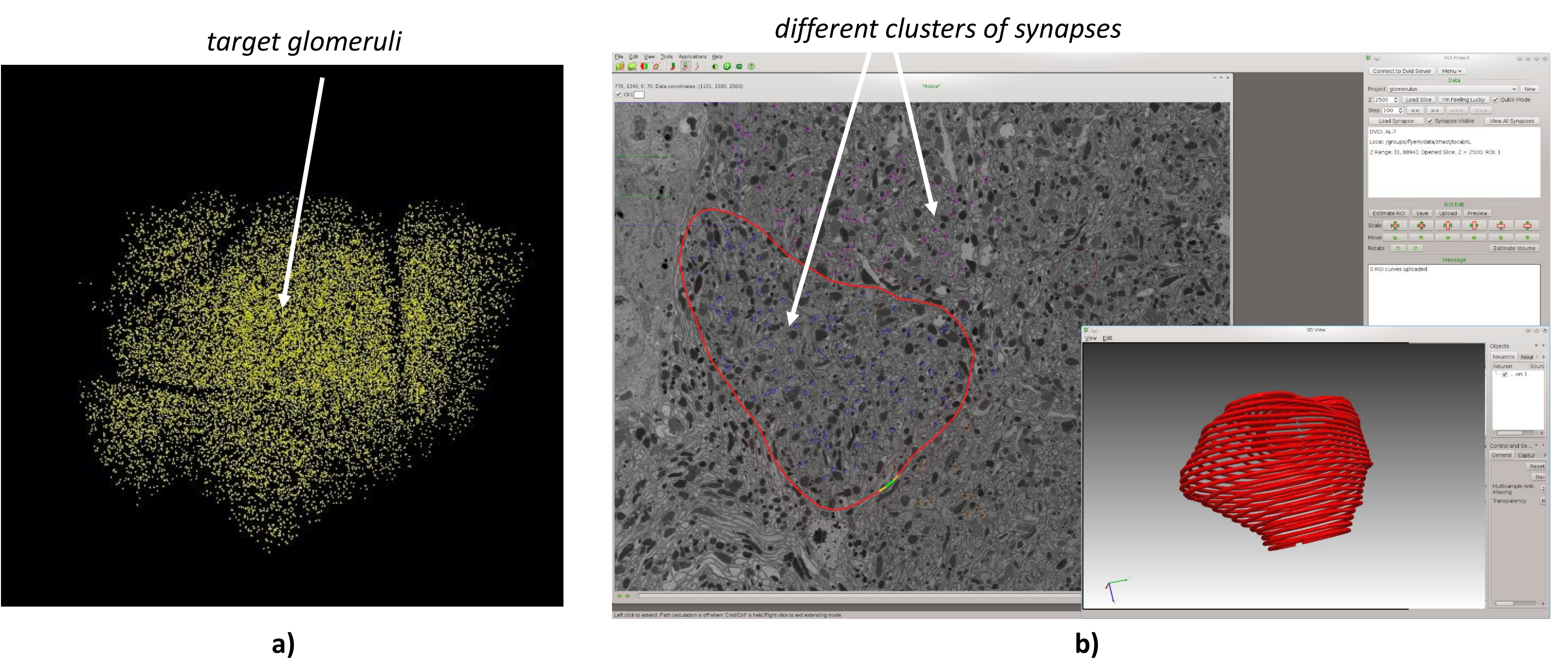}
\caption{\label{fig:alroimot} {\bf Determining Region of Interest to help focus proofreading efforts using synaptic information .}  a) Manually identified synapses in AL show that some synapses go beyond the target glomeruli (VA1v).  b)  Using these annotations, a specific ROI was determined using the tool NeuTu.}
\end{figure}

A total of $19921$ synapses were manually annotated taking a little over 38 8-hour
proofreader days
to complete.  These annotations were used as training for the synapse prediction but also indicated
the boundaries with adjacent glomeruli, as shown in Figure \ref{fig:alroimot}.a.  These synapses
were clustered into different regions.  By showing the pattern of their clusterings
and grayscale in NeuTu, we asked
an expert to define more precisely the glomerulus boundary by drawing loops through
the low-density synapse regions.

The resulting glomerulus ROI, identified to be a large portion of VA1v, is $4858$ $mu$m$^3$ compared with $11875$ $mu$m$^3$ of the original 
region manually annotated (the entire AL is over $285000$ $mu$m$^3$).
Only $11256$ synapses 
were actually located in glomerulus ROI.  If this ROI had been chosen before performing
any manual annotation, there would have been roughly a $43\%$ reduction in
the work/annotation.
Subsequent tracing and connnectome analysis can also be better defined with this ROI.  Note that the synapses per cubic micron manually annotated was $2.31$, close
to the $2.35$ listed for VL1 in Table \ref{tab:segstats}.  Note that the manually annotated and drawn ROI is actually still a subset of the whole VL1; a subset of the manual annotation was used to train the predictor.)

Our experiments also show that defining an ROI in NeuTu by manually drawing several 2D loops is time-efficient.  The glomerulus ROI and entire AL ROI required 45 and 84 loops respectively and took under a day to trace.

\section{Discussion}
We show that automatic analysis of large EM datasets,
such as uncovering different neuropil compartments, counting synapses, and estimating their packing density, is possible
using fully automatic if imperfect
synapse prediction protocols.  Sampling our predictions gives us confidence that
synapse counts are statistically accurate.

We believe automatic synapse prediction
could also be useful in guiding image segmentation and reducing manual effort in
connectome reconstruction.  For instance, eliminating manual synapse annotation in connectome reconstruction will
significantly reduce labor - the manual effort required to annotate
the entire AL dataset alone would require 1,000 proofreader days.  
Fortunately, in many cases and for many applications inaccuracy in
the prediction of synpatic sites might not harm downstream connectome
analysis.  First, many pathways have redundant
connections.  Also, note that in a previous report, the consistency
between manual annotators is only around $90\%$, not much better than with
our automatic approach \cite{plaza14synapse}.  (This consistency
could also be further
improved with more time-consuming, double-checking and consensus-based proofreading but for identification and characterization
of strong pathways, this would not in fact be required.)
Ultimately, using automatic synapse prediction in a connectome will require
more detailed analysis to ensure that subtle biases are not introduced.

We introduce synapse density statistics for the AL and several
constituent glomeruli.  We enlist the superior resolution of EM data
to make claims about the accuracy of our approach.  While small inaccuracies
in the synapse prediction currently prevent conclusive claims for differences in
synaptic density between glomeruli, both our manual annotation and predictions suggest
that a synapse packing density exists that is far greater than previously reported
using the construct Brp-short \cite{mosca14}.  
The authors of this previous report \cite{mosca14} reveal a 
consistent packing density of olfactory receptor neuron outputs
across several glomeruli in a 10-day adult {\em Drosophila}.
It is possible that Brp-short does not label all synapses, just as nc82 at a dilution used labels only a fraction of the T-bar ribbons in the optic lamina \cite{Meinertzhagen10}.
The differences in synaptic counts could also
be from AL interneurons or projection neurons.
However, specimen
differences complicate this comparison (for instance degeneration at synapses could already be present).
Future work could strengthen the analysis by generating predictions from multiple samples.  In addition, ongoing semi-automatic reconstruction efforts in the AL will eventually reveal the exact distributions of connections per neuron class.

We note that our synapse clouds are closely related to nc82 immunopuncta or other synapse labeled data.
For some analyses, EM synapse point clouds are a preferred alternative, for example
when synapse point labels require verification or when subclasses of synapses
need to be analyzed.  Another consideration is the unknown
effectiveness of nc82 in detecting synapses
with different compositions and maturities.  Brp is localized
to the platform
of the T-bar \cite{Fouquet09}.  In some EM data, we note that the platforms are not fully developed.  It is
unclear how these T-bars will be labeled in an nc82 sample, whereas, a more detailed
analysis could be undertaken in the EM data.  Anecdotally, we observe that our synapse cloud
contains these partial synapse sites.

The biggest obstacle to high-throughput generation of  EM synapse point clouds is the speed of image acquisition.  Synapse prediction requires minimal training and can be computed quickly on a cluster.  Continual
advances in imaging technology promise to
reduce acquisition bottlenecks greatly.

{\small
\textbf{Acknowledgements:} We thank Zhiyuan Lu for sample preparation; Shan Xu and Harald Hess for FIB-SEM imaging; 
the FlyEM proofreading team (Lei-Ann Chang and the Dalhousie proofreading team) for the annotations efforts; and Yang Yu
for EM/light registration.}

\end{document}